\documentclass[graybox]{svmult}

\usepackage{helvet}         
\usepackage{courier}        
\usepackage{type1cm}        
\usepackage{amsmath}
\usepackage{makeidx}         
\usepackage{graphicx}        
\usepackage{multicol}        
\usepackage[bottom]{footmisc}

\begin{document}

\title*{Entanglement and the Infrared}
\author{Gordon W. Semenoff}
\institute{Gordon W. Semenoff \at Department of Physics and Astronomy, University of British Columbia, 6224 Agricultural Road, Vancouver, British Columbia, V6T 1Z1 Canada, \email{gordonws@phas.ubc.ca}
}
\maketitle

\abstract{We shall outline some results regarding the infrared catastrophes of quantum electrodynamics and perturbative quantum gravity and their implications for information loss in quantum processes involving electrically or gravitationally charged particles.
We will argue that two common approaches to the solution of the infrared problem, using transition probabilities which are inclusive of copious soft photon and graviton production and using dressed states describe fundamentally different quantizations 
of electrodynamics and low energy gravity which are, in principle, distinguishable by experiments.
}

\section{Prologue}

Motivated by the idea that subtle infrared effects could be relevant to the black hole information paradox, interest in the infrared problems in quantum electrodynamics and in perturbative quantum gravity has recently seen a rebirth 
\cite{Ware:2013zja}-\cite{Hannesdottir:2019opa}. These happen to be the two known theories of nature which contain massless physical particles and which describe long-ranged interactions. 
There are two well developed ways of dealing with the infrared divergences in these theories.
 
 The first of the two has been known since the early days of quantum electrodynamics \cite{Bloch:1937pw}-\cite{Jauch:1976ava}, and was generalized to perturbative quantum gravity by Weinberg \cite{Weinberg:1965nx}.  In this  approach, the infrared divergences that occur in internal loops in Feynman diagrams, and which afflict the $S$-matrix that is computed in renormalized perturbation theory, are canceled by computing the probabilities of processes which also include the production of  soft photons and soft gravitons.  In this approach, the infrared divergences of the perturbative $S$-matrix cancel with those which occur in the integration of transition probabilities over the wave-vectors of the outgoing soft particles, leaving infrared finite inclusive transition probabilities.  The precise order by order cancellation of the infrared divergences by this mechanism is due to unitarity and it can be seen as a consequence an optical theorem for the S-matrix. 
 
The second formalism considers dressed states where the quantum states of charged particles are dressed by adding soft on-shell photons and gravitons. The soft particle content of the dressed state is fine-tuned in such a way that transition amplitudes between dressed states are infrared finite \cite{Chung:1965zza}-\cite{Kulish:1970ut}.  Moreover, to an accuracy which is governed by the detector resolution, the transition probabilities which are computed in this second approach are identical to those of the first approach.  

The replacement of charged particle states by dressed states can be implemented as a canonical transformation \cite{Javadinazhed:2018mle} which decouples the infrared, so that the copious production of arbitrarily soft particles, beyond those already included in the dressed states, no longer occurs in a scattering processes.  In this approach, the $S$-matrix elements between dressed states is infrared finite. However, the canonical transformation which dresses the charged particles  is an improper unitary transformation. All of the dressed states are orthogonal to all of the multi-particle Fock states. As a result, the first and second approaches are not equivalent, they have different, orthogonal, Hilbert spaces. They should be considered different, inequivalent theories of how to deal with infrared divergences.   

Recently, it has been noted that the two approaches, dressed and un-dressed, have important and potentially physically observable differences in how  quantum information is distributed by the interactions when a scattering process occurs \cite{Carney:2017jut}-\cite{Neuenfeld:2018fdw}.  It is known that even elastic scattering results in entanglement of the quantum states of the out-going particles \cite{Park:2014hya}-\cite{Grignani:2016igg}. In the first approach to the infrared, the copious production of a cloud of soft photons or soft gravitons, which then fly away, undetected, from a scattering event, results in a quantum state where the  soft photon or soft graviton cloud and the hard particles that are left behind are highly entangled. The result of this entanglement and the inaccessibility of the soft photon cloud to measurements  is decoherence which, although very small in any realistic experiment, could in principle be measured.  If the particles are dressed, and the infrared is decoupled, so that pure states evolve to pure states, this fundamental decoherence must be absent. 

We will mostly use the language of quantum electrodynamics in the following as we anticipate that it may be more familiar to the reader.  Practically all of our considerations also apply to perturbative quantum gravity in the low energy regime and we will give some of the relevant formulae.  Of course quantized gravity is not a consistent, renormalizable quantum field theory.  Moreover, it is not clear that it can have an infrared cutoff which leaves it unitary. We will ignore these difficulties here. 

\section{Inclusive approach to infrared singularity cancellation}

If we wanted to use quantum electrodynamics to compute the amplitude for Moller scattering, for example, we would begin with the Feynman diagram which is illustrated in figure 1.  That diagram  gives an estimate of the quantum  amplitude that  two incoming  electrons  will interact and then re-emerge as two electrons.  The modulus square of this amplitude  gives an answer for the probability that the process will happen which, because of the small value of the electromagnetic coupling constant $\frac{e^2}{4\pi}\sim\frac{1}{137}$ is already accurate to one percent.  

 \begin{figure}
 ~~~~~~~~~~~~~~~~~~~~~~~~~~~~~~~ \includegraphics[scale=.11]{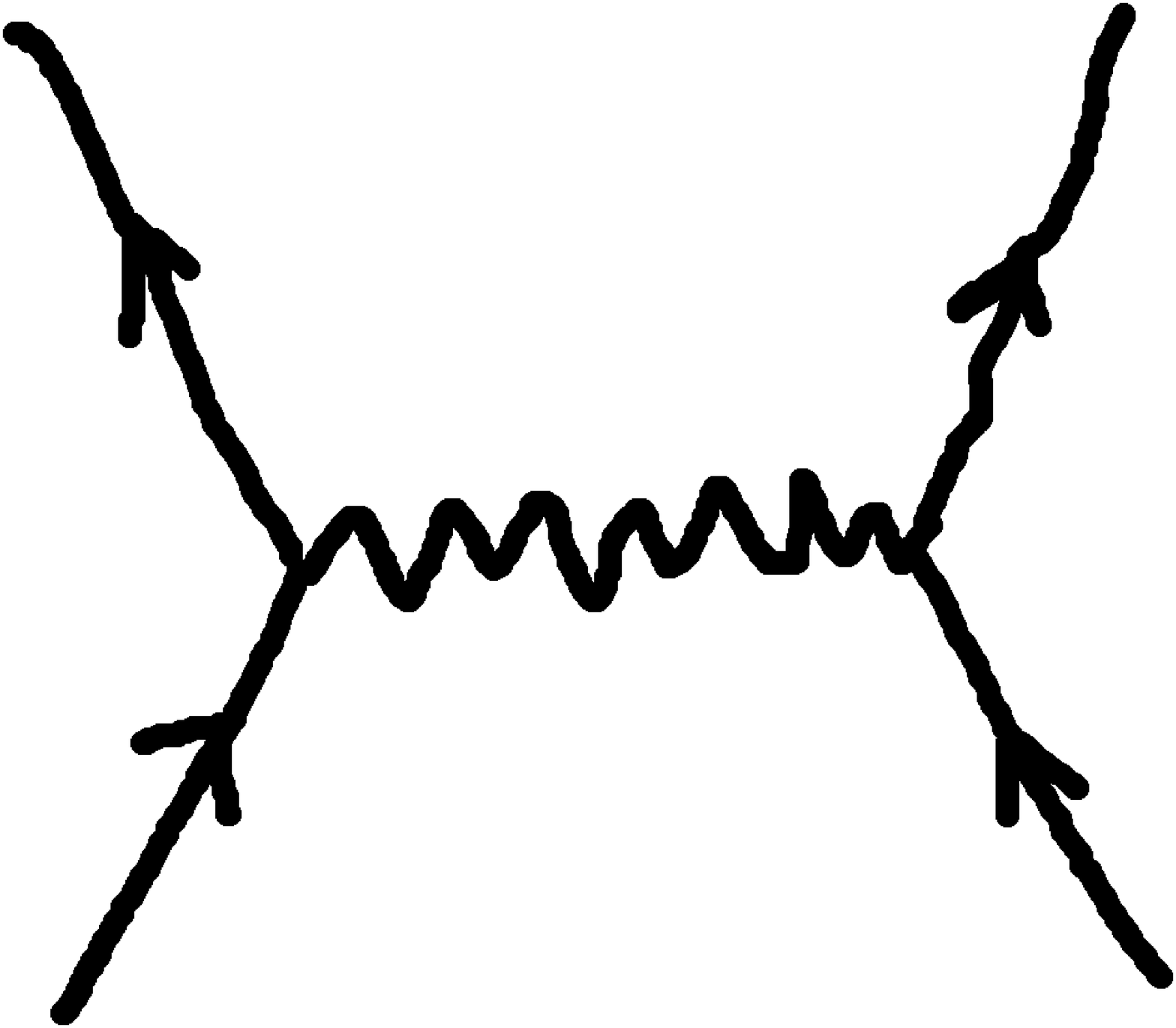}
\begin{caption}{The Feynman diagram which is used to compute the quantum amplitude for Moller scattering is depicted. The probability of the two-electron state, incoming from the bottom of the diagram, emerging as a two-electron state is gotten by taking the square of the modulus of this amplitude.  Because quantum electrodynamics is a weakly coupled theory, the result is already accurate to the one percent level. }\end{caption}
\label{moller1}\end{figure}

If we want to improve the accuracy of the computation, we must include higher order corrections in the way of loop diagrams.  The next correction occurs at one loop and it consists of several processes.  One of them is illustrated in the second diagram in figure 2 where the electron emits a virtual photon, interacts with the other electron
and then re-absorbs the  virtual photon.  This contribution will be infrared divergent.  Unlike ultraviolet divergences, which are well understood, and are dealt with by using the usual renormalization procedure, the infrared divergence is physical and it must be dealt with by using physical reasoning.

 \begin{figure}
~~~~~~~~~~~~~~~~~~  \includegraphics[scale=.3]{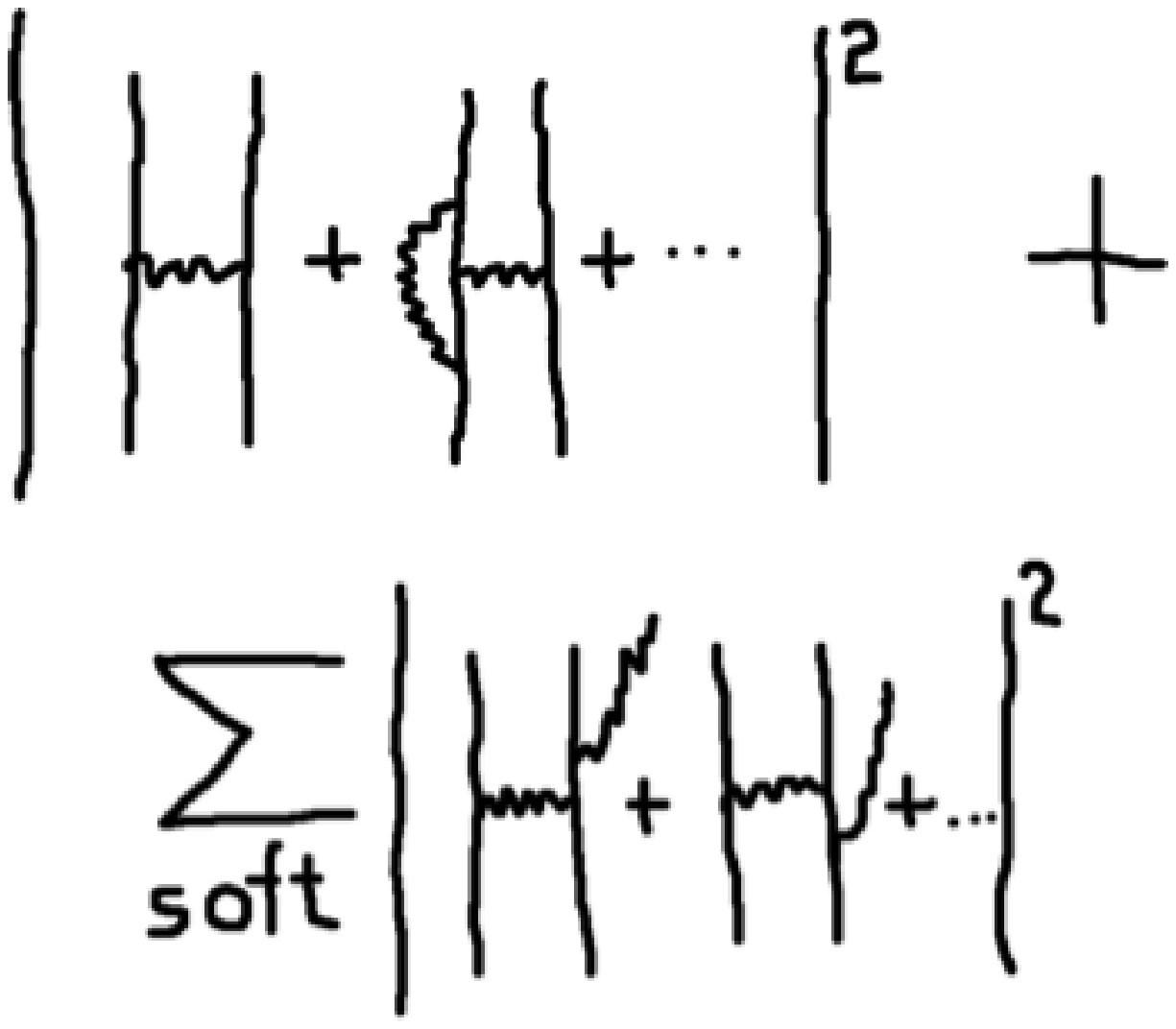}
\begin{caption}{The probability of a Moller scattering process is gotten by taking the squared modulus of the sum of the leading order and higher order Feynman diagrams which contribute to Moller scattering amplitude and then adding a similar squared modulus of the amplitude for Moller scattering plus the production of a soft photon.  Here, only one example diagram of the several that contribute at the next-to-leading order are displayed.  The infrared divergence from the internal loop is canceled by the integration over soft momenta of the extra emitted photon.  The cancelation is between the last term and the cross term in the first bracket. }\end{caption}
\label{moller2}\end{figure}

The solution to this infrared problem is well known and it dates back to the early days of quantum electrodynamics  \cite{Bloch:1937pw}-\cite{Jauch:1976ava}, in fact it predates the understanding of ultraviolet renormalization by a few decades.  The solution is to consider an additional process which is physically indistinguishable from the process that we have described up to now.  That process considers the same Moller scattering, but with the additional production of a soft photon.  The photon should be so soft that it
eludes detection by the detection apparatus, and thus, it flies away undetected from our Moller scattering experiment.  The idea is that we should add the possibility of this process to the one where no soft photon is produced.  That probability is the one represented by the last term in figure 2. If that last contribution is integrated over the wave-vectors of the soft photon, it is also infrared divergent.  In fact, it is divergent in such a way as to cancel the infrared divergence in the same order ($e^6$) cross-term in the first contribution.  This cancellation is exact.  Its fine-tuning  
is a result of unitarity -- the optical theorem -- and this sort of argument can be seen to cancel the infrared divergences encountered in any amplitude which involves charged particle scattering and to all orders in perturbation theory.  

An important consequence of the argument in the paragraphs above is the fact that, even though the lowest order Feynman diagram in figure 1 turns out to be the correct one to accurately analyze Moller scattering,  the physics of what is happening is much more complicated.  The relative amplitude for the process in figure 1 is zero.  The processes which dominate are those where infinite numbers of soft photons
are produced, as in figure 3.   
   \begin{figure}
 ~~~~~~~~~~~~~~~~~~~~~~~~~~~~~~~ \includegraphics[scale=.08]{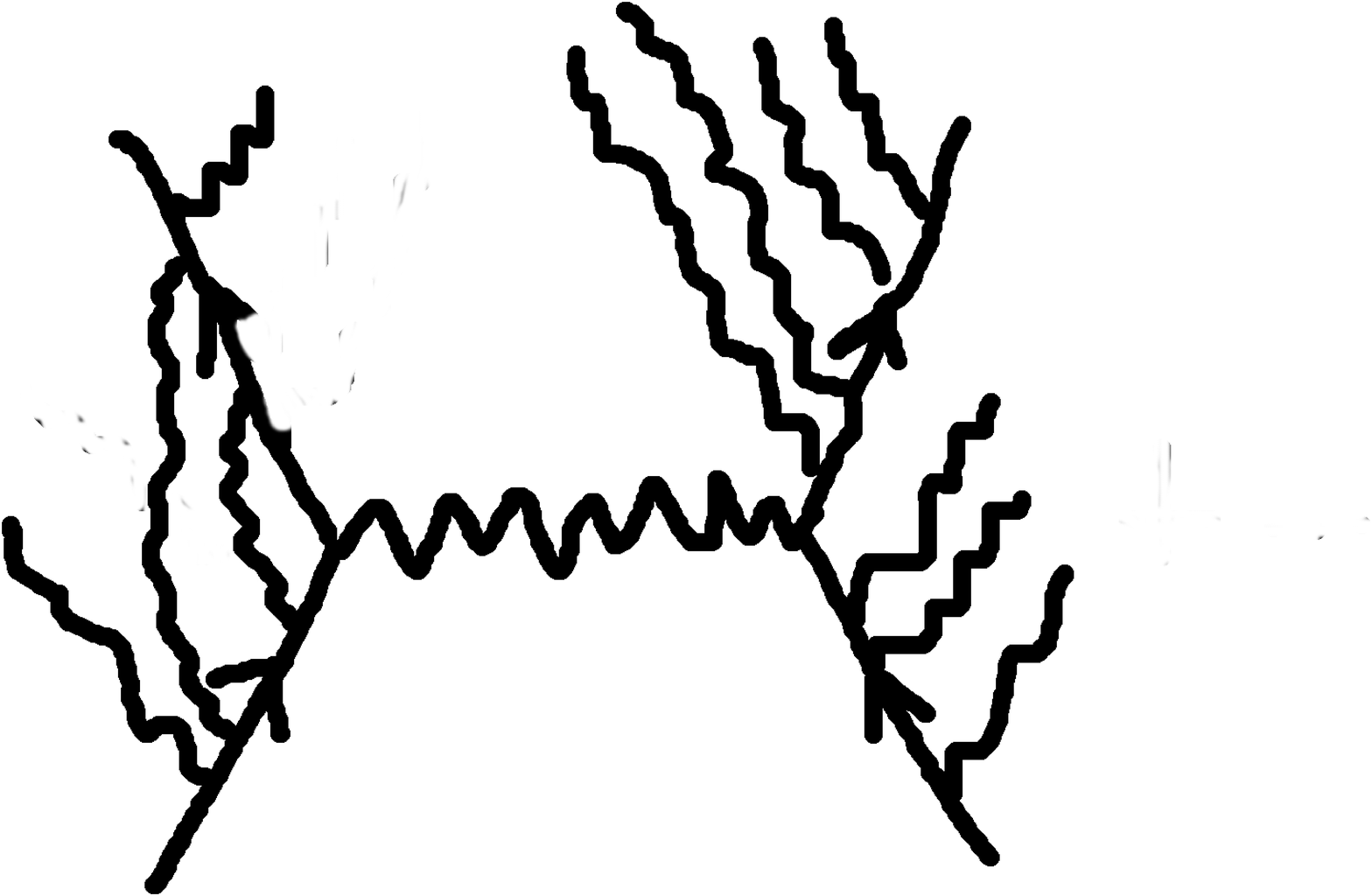}
\begin{caption}{The physical processes which contribute to Moller scattering and which 
have non-zero probability involve the copious production of soft photons. }\end{caption}
\label{moller3}\end{figure}
These are needed to cancel the divergences in internal loops. 
  
 The infinite numbers of photons which fly away undetected carry very little energy or momentum. To the accuracy of the detector resolution, their influence on the kinematics of the experiment is not noticeable.  However, even if they have very little energy,  each photon has a polarization and a direction of motion.  Specifying the details of their quantum state involves a significant amount of information.  A question that one could then ask is, when this cloud of photons escapes detection, how much information is lost?  What we mean here  is information in the quantum sense, as we shall try to explain in the next section. 
 This question has only been recently addressed \cite{Carney:2017jut}-\cite{Neuenfeld:2018fdw} and as we will explain in the rest of this review, the results were somewhat surprising.

  \section{Information loss due to quantum entanglement}

Let us try to explain precisely what we mean by information loss.   Let us consider a model system of two qubits, qubit $\#1$ and qubit $\#2$.  We could think of qubit $\#1$ as the analog of the hard particles in our scattering experiment and quit $\#2$ as the soft photons and gravitons that are produced.  Bases for the Hilbert spaces of the quantum states of qubit $\#1$ are the two vectors $|\uparrow>_1$ and $|\downarrow>_1$ and for the qubit $\#2$ the states
$|\uparrow>_2$ and $|\downarrow>_2$.  Let us assume that the qubits are dynamically independent, that is, they do not interact with each other.  

The question that we want to ask is, if qubit $\#2$ becomes inaccessible to us, how much information about the quantum state of qubit $\#1$ have we lost.  In the classical world, if these were classical bits, rather then qubits, the answer would be easy -- none!  Everything that we could find out by classical measurements of qubit $\#1$ before qubit $\#2$ was misplaced could still be done afterward.  As far as qubit $\#1$ is concerned, we would have lost no information at all.  

In the quantum world 
the answer will depend on the quantum state of the joint two-qubit system at the time when qubit $\#2$ was lost.  
Let us consider two examples for that quantum state, an un-entangled state
$$
|\psi> =\left[\cos\varphi |\uparrow>_1+\sin\varphi|\downarrow>_1\right]\otimes|\uparrow>_2
$$
and an entangled state
$$
|\tilde\psi> = \left[\cos\varphi |\uparrow>_1\otimes|\uparrow>_1 +\sin\varphi |\downarrow>_1\otimes|\downarrow>_2\right]
$$
  
These two states have the same expectation values of the ``spin'' of qubit 1, that is, the expectation values of 
the operator $|\uparrow>_1<\uparrow|\otimes{\cal I}_2$, which is $\cos^2\varphi$ or $|\downarrow>_1<\downarrow|\otimes{\cal I}_2$ which is $\sin^2\varphi$. 
The difference between the two states is that the un-entangled state $|\psi>$ has a wave-function which is a direct product of the wave-functions of qubit $\#1$ and qubit $\#2$.  The entangled state, $|\tilde\psi>$, on the other hand, is a superposition of direct products, which cannot itself be written as a single direct product of states of $\#1$ and $\#2$.   

Now, let us assume that, in the quantum world, we have lost track of qubit $\#2$.  What is the implication  for qubit $\#1$.
We get information from a quantum system by quantum measurements.  Quantum measurements are represented mathematically by projection operators.  If we have no access to qubit $\#2$, all of the quantum measurements that we can do must act on  qubit $\#2$ like the unit operator on its factor in the Hilbert space.  Therefore, for the sake of quantum measurements, we can once and for all contract the states of qubit $\#2$ with the unit operator, that is, we can form the reduced density matrix which describes qubit $\#1$, in our first example, by tracing over the states of qubit $\#2$,
$$
\rho = {\rm Tr}_2~ |\psi><\psi| = \left[~ \cos\varphi |\uparrow>_1+\sin\varphi|\downarrow>_1~\right]\left[~_1<\uparrow|\cos\varphi
+_1<\downarrow|\sin\varphi~\right]
$$
$$
=\left[ \begin{matrix} \cos^2\varphi & \cos\varphi\sin\varphi\cr 
\cos\varphi\sin\varphi &\sin\varphi\sin\varphi \cr \end{matrix}\right]
$$
or, in our second example,
$$
\tilde\rho ={\rm Tr}_2~ |\tilde\psi><\tilde\psi| =\left(\cos^2\varphi  |\uparrow>_1<\uparrow| +
\sin^2\varphi|\downarrow>_1<\downarrow|\right)
$$
$$
 =\left[ \begin{matrix}\cos^2\varphi & 0\cr 0 & \sin^2\varphi \cr \end{matrix}\right]
$$
In the first, unentangled example, the reduced density matrix is still that of a pure state.  Qubit $\#1$ is sure
to be in the quantum state $ \left[ \cos\varphi |\uparrow>_1+\sin\varphi|\downarrow>_1\right]$.  No information about its state has been lost.  However, in the second case, the reduced density matrix is now that of a mixed state with classical probabilities $\cos^2\varphi$ of finding $|\uparrow>_1$ and $\sin^2\varphi$ of $|\downarrow>_1$.  What is missing are the off-diagonal elements of the density matrix. These contain interference terms.  We can see the difference if we ask what is the expectation value of quit $\# 1$ is in a state which is a superposition of spin up and spin down, $\left[\alpha|\uparrow>_1+\sqrt{1-|\alpha|^2}|\downarrow>_1\right]$, which is gotten by tracing the reduced density matrix times
the projection operator 
$$
{\mathcal O}= \left[\alpha|\uparrow>_1+\sqrt{1-|\alpha|^2}|\downarrow>_1\right]
\left[ ~_1<\uparrow|\alpha^*+~_1<\downarrow|\sqrt{1-|\alpha|^2}\right]
\otimes{\cal I}_2
$$   
In the first case, the expectation value is 
$$
{\rm Tr}{\mathcal O}\rho=\left| \cos\varphi~\alpha+\sin\varphi\sqrt{1-|\alpha|^2}\right|^2
$$
whereas in the second case it is
$$
{\rm Tr}{\mathcal O}\tilde \rho=\left| \cos\varphi~\alpha\right|^2+\left|\sin\varphi\sqrt{1-|\alpha|^2}\right|^2
$$
The difference is, in the second case the cross-terms, that is, the interference terms are missing.  In the second case, we have lost the possibility of interference.  This is called decoherence.   In the entangled case, when qubit $\#2$ was lost, the quantum probabilities of the two spin outcomes became classical probabilities.  On the other hand, in the un-entangled case, no information was lost.  The outcomes of all possible measurements of qubit $\#1$ remain unchanged.  

The property of the state $|\tilde\psi>$ which distinguishes it from state $|\psi>$ and which results in decoherence is quantum entanglement.  
A quantitative measure of entanglement is the entanglement entropy, defined as the Von Neumann entropy of the 
reduced density matrix,
$$
S = -{\rm Tr} \rho\ln\rho
$$
In the un-entangled case, $S=0$, whereas in the entangled case, $\tilde S=-\cos^2\varphi\ln\cos^2\varphi-\sin^2\varphi\ln\sin^2\varphi$.

\section{Entanglement of soft and hard}

Lets us return to quantum electrodynamics and consider a scattering event where an incoming state $|\alpha>$  evolves to an out-going state.   
The outgoing state is a superposition of incoming states.   The coefficients in this super-position are the elements of the $S$-matrix,
$$
|\alpha>~\to~ \sum_{\beta,\gamma}|\beta,\gamma>S_{\beta\gamma,\alpha}^{\dagger}
$$
Here, in  $|\beta,\gamma>$, we are separating the soft photons, which we call $\gamma$, from the hard particles, which we denote by $\beta$.

In a perturbative computation, the $S$ matrix turns out to be  infrared divergent and an infrared cutoff
is needed in order to define it.  We shall introduce such an infrared cutoff which we will denote by $\mu$.  
A nice example of how this could be done is to assume that the photon has a small mass, $\mu$, so the Maxwell theory coupled to charged matter becomes the Proca theory of a massive vector field, coupled to
the conserved charged currents of the charged matter.  This is still a Lorentz invariant, renormalizable quantum field theory
with a unitary $S$-matrix that we shall denote $S_{\alpha\beta}^\mu$ where the superscript $\mu$ reminds us that it is to be computed with the infrared cutoff $\mu$ in internal loops.  The infrared cutoff $S$-matrix is unitary,
$$
\sum_\alpha S^{\mu\dagger}_{\beta\gamma\alpha}S^\mu_{\alpha\beta'\gamma'}= \delta_{\beta\beta'}\delta_{\gamma\gamma'}
$$
where the sum on the left-hand side is schematic for integrations and sums over the momenta and quantum numbers 
of the particles in the incoming state and the right-hand-side is schematic for an assembly of Dirac and Kronecker delta functions which identify momenta and discrete quantum numbers in the states $|\beta,\gamma>$
and $|\beta',\gamma'>$.  

Generally, our incoming states can be either eigenstates of energy and momentum or they can be wave-packets.
In order to address the most general consideration, we will consider an in-coming density matrix of the form
$$
 |\alpha><\alpha'|
$$
where $\alpha$ and $\alpha'$ are states where each of the incoming particles has a fixed energy and momentum. 
If these states contain photons, they are hard photons, with energies and wave-vectors much larger than the fundamental
infrared cutoff $\mu$ and we will also need them to be much larger than another intermediate cutoff, which we shall call $\lambda$, the detector resolution. 

We could make a wave-packet incoming state from this incoming density matrix by taking the superpositions
$$
|f><f|~\equiv~
\sum_{\alpha\alpha'}f_{\rm in}(\alpha)f_{\rm in}^*(\alpha')|\alpha><\alpha'|
$$
and this state is normalized if
$$
\sum_\alpha |f_{\rm in}(\alpha)|^2=1~,~<f|f>=1
$$

During the scattering process, the in-state $|\alpha><\alpha'|$  evolves to  specific out-state which is given by
$$
|\alpha><\alpha'|~\to~\left[\sum_{\beta\gamma}|\beta\gamma>S^{\mu\dagger}_{\beta\gamma\alpha}\right] \left[\sum_{\beta'\gamma'}S^\mu_{\alpha'\beta'\gamma'} <\beta'\gamma'|\right]
$$
where we have separated the scattering products into hard particles, $\beta,\beta'$, those whose momenta are above
a the detector resolution cutoff,  $\lambda$,  and the soft particles $\gamma,\gamma'$ whose
frequencies and wave-numbers are smaller than the detector resolution $\lambda$ but greater than the fundamental cutoff,$\mu$.  Any state of free particles can be divided in this way.

We then reduce the density matrix of the final state by tracing over the soft degrees of freedom.  This yields
$$
\rho =\sum_{\mu<\tilde \gamma<\lambda}<\tilde\gamma|\rho_{\rm out} |\gamma>
=\sum_{\mu<\tilde \gamma<\lambda}<\tilde\gamma| \left[\sum_{\beta\gamma}|\beta\gamma>S^{\mu\dagger}_{\beta\gamma\alpha}\right] \left[\sum_{\beta'\gamma'}S^\mu_{\alpha'\beta'\gamma'} <\beta'\gamma'|\right]
|\tilde\gamma> 
$$
or, simplifying the notation, 
\begin{equation}\label{shored}
<\beta|\rho |\beta'>=  \sum_{\mu<  \gamma<\lambda}
S^{\mu *}_{\alpha\beta\gamma }S^\mu_{\alpha'\beta'\gamma}
\end{equation}
Now, we would like to use a soft photon theorem to simplify this expression, particularly the trace over soft photons.  A nice derivation and discussion of the soft photon theorem can be found in Weinberg's quantum field theory book \cite{Weinberg:1995mt}. 

 A soft photon theorem is valid only when we have a large hierarchy of scales.  That means that we can apply it to our out-state only when the masses, energies and momenta of all of the particles in the states $|\beta>$ and $|\beta'>$
are much greater than the detector resolution, $\lambda$, and also when $\lambda$ is much greater than the fundamental cutoff, $\mu$. 
This means that we cannot analyze every possible out-state, but only those which meet this requirement.  We will not worry about this limitation here or in the following.  We emphasize that shall also need that $\lambda>>\mu$.  In addition, for completeness, 
we shall cut off the total energy of the photons that escape with a cutoff $E$.  To be clear, $E$ is the maximum value
of the sum of all of the energies of the soft photons.   
The soft photon theorem then tells us that, when there is a hierarchy of scales, $E,\lambda>>\mu$, we can replace
equation (\ref{shored}) by the expression 
\begin{align}
&<\beta|\rho |\beta'>=
S^{\mu\dagger}_{\beta\alpha}S^\mu_{\alpha'\beta'}
\left(\frac{\lambda}{\mu}\right)^{A_{\alpha\beta,\alpha'\beta'}}f\left( \frac{\lambda}{E},A_{\alpha\beta,\alpha'\beta'}
\right)
\end{align}
where the exponent is a complicated function of the four-momenta of the hard particles in the initial and final states,
\begin{align}\label{A}
&A_{X,Y} = - \sum_{n\in X}\sum_{n'\in Y}
\frac{e_ne_{n'}\eta_n\eta_{n'}}{8\pi^2} \beta_{nn'}^{-1}
\ln \frac{1+\beta_{nn'} }{1-\beta_{nn'} }
\end{align}
where  $e_n$ are the charges of particles, and $\eta_n=1$ for an
incoming particle and $\eta_n=-1$ for an outgoing particle. 
$$
\beta_{nn'}=\sqrt{1-\frac{(m_nm_{n'})^2}{(p_n\cdot p_{n'})^2}}
$$ 
are the relativistic relative velocities of particles $n$ and $n'$. 
The last factor comes from imposing the cutoff on the total anergy and it contributes
\begin{align}
&f(\lambda/E,A)=\frac{1}{\pi}\int_{-\infty}^\infty du\frac{\sin u}{u}\exp\left(A\int_0^{\lambda/E}\frac{d\omega}{\omega}\left( e^{i\omega u}-1\right) \right)
\\& f(1,A)=
\frac{ e^{-\gamma A} }{\Gamma[1+A]}~,~\gamma=.05772...~,~~f(0,A)=1
\end{align}
We have included the result with the  energy cutoff $E$ for completeness, however, it will play no role in 
the following, so we will put $E\to\infty$ where $f(\lambda/E,A)\to 1$.  

The trace over soft photons produces energy and momentum-dependent factors
multiplying the $S$-matrix for the hard particles alone. These factors, as well as the
$S$-matrix, depend on the fundamental cutoff $\mu$.  Now that we have assumed a hierarchy
of scales, we can also exchange the infrared cutoff $\mu$ for a larger cutoff $\Lambda$ where it appears
in the $S$-matrix.  We can choose the new cutoff $\Lambda$ and 
  a further soft photon theorem tells us that
\begin{align}\label{soft2}
S^\mu_{\alpha\beta} = S^\Lambda_{\alpha\beta}\left( \frac{\mu}{\Lambda}\right)^{A_{\alpha\beta,\alpha\beta}/2}
\end{align}
The right-hand-side of this equation does not depend on $\Lambda$, at least over a range of $\Lambda$ that
respects the hierarchy of scales
$
\alpha\alpha'\beta\beta' >>E,\lambda,\Lambda>>\mu
$,
where, by $\alpha\alpha'\beta\beta'$, we mean the energies  of all of the particles in the states labeled by  $\alpha\alpha'\beta\beta'$. 
Using this equation, we find
\begin{align}\label{rho_red}
&<\beta|\rho|\beta'>= 
S^{\Lambda *}_{\alpha\beta} ~
S^\Lambda_{\alpha'\beta'}~\left( \frac{\mu}{\Lambda}\right)^{A_{\alpha\beta,\alpha\beta}/2}
\left( \frac{\mu}{\Lambda}\right)^{A_{\alpha'\beta',\alpha'\beta'}/2} ~
\left(\frac{\lambda}{\mu}\right)^{A_{\alpha\alpha',\beta\beta'}}
\end{align}
This is our expression for the reduced density matrix that we use to describe the quantum state of the out-going hard
particles. We can see, by studying the exponents, $A_{XY}$ that its diagonal matrix elements, for a fixed in-state $\alpha=\alpha'$,   are
\begin{align}\label{diagonal}
&<\beta|\rho|\beta>= 
|S^\Lambda_{\alpha\beta}|^2~
\left(\frac{\lambda}{\Lambda}\right)^{A_{\alpha \beta,\alpha\beta}}
\end{align}
which no longer depends on the fundamental cutoff.  In fact it simply has the form of the square of the transition amplitude for $|\alpha>\to |\beta>$, computed with an infrared cutoff $\Lambda$ for internal loops in Feynman diagrams, 
 times the Sudakov-like factor $\left(\frac{\lambda}{\Lambda}\right)^{A_{\alpha \beta,\alpha\beta}}
$ with the ratio of detector resolution $\lambda$ and the infrared cutoff $\Lambda$.  This result is well known.  
Note that, due to equation (\ref{soft2}), equation (\ref{diagonal}) actually does not depend on $\Lambda$.  

Now, what about the off-diagonal elements? They can be written as
\begin{align}\label{rho_red}
&<\beta|\rho|\beta'>= 
S^{\Lambda * }_{\alpha\beta} ~
S^\Lambda_{\alpha'\beta'}
\left(\frac{\lambda}{\Lambda}\right)^{A_{\alpha\alpha',\beta\beta'}}
~\left( \frac{\mu}{\Lambda}\right)^{\Delta A(\alpha\alpha'\beta\beta')}
\end{align}
Now the small $\mu$ behaviour is dependent on the exponent
$$
\Delta A(\alpha\alpha',\beta\beta' )= A_{\alpha\beta,\alpha'\beta'}-A_{\alpha\beta,\alpha\beta}/2-A_{\alpha'\beta',\alpha'\beta'}/2\geq 0
$$
This exponent can be shown to be positive semi-definite \cite{Carney:2017jut},
\cite{Carney:2018ygh}
. This means that, as we remove the fundamental
cutoff, to make the photon truly massless, some off-diagonal elements of the density matrix are set to zero.  Only
those where $\Delta A(\alpha\alpha',\beta\beta')=0$ survive.  This turns out to be a surprisingly strict restriction on
which elements survive.  It turns out that, 
$\Delta A(\alpha\alpha',\beta\beta')=0$ if and only if the four sets of ingoing and outgoing electric currents obey
$$
\left\{ \frac{e_i p_i^\mu}{\sqrt{\vec p_i^2+m^2}} : e_i,p_i\in \alpha\right\}
=\left\{ \frac{ e_jp_j^\mu}{\sqrt{\vec p_j^2+m^2}} : e_j,p_j \in \beta'\right\}
$$
$$
\left\{\frac{ e_k p_k^\mu}{\sqrt{\vec p_k^2+m^2}} : e_k,p_k\in \alpha'\right\}
=\left\{ \frac{e_\ell p_\ell^\mu}{\sqrt{\vec p_\ell^2+m^2}} : e_\ell,p_\ell\in \beta\right\}
$$
where the equalities are up to permutations of the elements. 
That is, the sets of electric currents are identical.
In conclusion,
the matrix element of the reduced density matrix survives if and only if the set of all electric currents
contained in the states $\alpha,\beta'$ is identical (up to permutations) to the set of all electric currents
in the states $\alpha',\beta$.  If these currents do not match, $\Delta A>0$ and the matrix elements vanish in 
the limit where the photon is massless.
Perturbative quantum gravity has a similar conclusion with the matching condition on the in-coming and outgoing energy-momentum currents
$$
\left\{ \frac{p^\nu_i p_i^\mu}{\sqrt{\vec p_i^2+m^2}} : e_i,p_i\in \alpha\right\}
=\left\{ \frac{ p^\nu_jp_j^\mu}{\sqrt{\vec p_j^2+m^2}} : e_j,p_j \in \beta'\right\}
$$
$$
\left\{\frac{ p^\nu_k p_k^\mu}{\sqrt{\vec p_k^2+m^2}} : e_k,p_k\in \alpha'\right\}
=\left\{ \frac{p^\nu_\ell p_\ell^\mu}{\sqrt{\vec p_\ell^2+m^2}} : e_\ell,p_\ell\in \beta\right\}
$$
This is remarkably restrictive.  If we assume that the incoming state is a pure state with in-coming plane waves, $\alpha=\alpha'$, we find that the off-diagonal elements of the density matrix vanish unless the electric and energy-momentum currents in the two states match exactly.  For some simple processes, this can mean that the out-going density matrix has diagonal blocks.  Of course this argument says nothing about the  diagonal elements themselves, they are as they have always been, the transition probabilities between plane-wave states  given in equation (\ref{diagonal}).  

The zeroing of off-diagonal elements of the density matrix is decoherence.  One loses the quantum coherence that is necessary for quantum interference to occur. An even more dramatic effect occurs with incoming wave-packets, superpositions of plane-wave states.  There, scattering seems to be suppressed in many cases. For example, if we look at even diagonal components of the 
final state density matrix for in-coming wave-packets,
$$
<\beta|\rho|\beta>=\sum_{ij}f(\alpha_i)f^*(\alpha_j)S^{\Lambda*}_{\alpha_i\beta} ~
S^\Lambda_{\alpha_j\beta}
\left(\frac{\lambda}{\Lambda}\right)^{A_{\alpha_i\alpha_j,\beta\beta}}
~\left( \frac{\mu}{\Lambda}\right)^{\Delta A(\alpha_i\alpha_j\beta\beta)}
     $$
     and the massless limit of the photon still requires that we now put $\mu\to 0$.  This, at least partially, concentrates the 
     sum over $\alpha_i,\alpha_j$ in the region $\alpha_i\to\alpha_j$.  However, the limit $\alpha_i\to\alpha_j$ is generally finite and the sums are actually integrals which then do not have enough support to get a non-zero result.  
   In this way,  the limit $\mu\to 0$  
     suppresses scattering of wave-packets.  There are many processes for which only the unit matrix part of the S-matrix will contribute to the scattering of wave packets \cite{Carney:2018ygh}.  
     
     \section{Dressed quantum states}
     
     Now we turn to the second way of dealing with the infrared problem, that of dressing the incoming and out-going states of charged particles with soft on-shell photons and gravitons with the dressing fine tuned in a way that cancels the infrared singularities.  For a given distribution of incoming currents, the dressed state is obtained by a canonical transformation which creates a coherent state of the photons which is tuned to the currents of the charged particles, 
\begin{align}
&| p_1,p_2,...>
\to|p_1,p_2,...>_D
\nonumber\\
&\equiv \exp\left(-e\int^{\lambda}_{\mu} \frac{d^3k}{\sqrt{2|k|}} \sum_{j\in \alpha}\frac{ p^{\mu}_j\epsilon^s_\mu(k)}{p^{\nu}_jk_\nu+i\epsilon} a_s(k)-  ~{\rm h.c. }  \right)| p_1,p_2,...>
\label{dress1}\\ &
a_s(k)\to  a_s(k)+  \frac{e}{\sqrt{2|k|}}  \sum_{j}\frac{ p^{\mu}_j\epsilon^s_\mu(k)}{p^{\nu}_jk_\nu-i\epsilon} ~,~
\label{dress2}
\mu<|\vec k|<\lambda
\nonumber
\end{align}     
where $\epsilon^s(k)$ is the physical polarization of the photon.
 If we take matrix elements of the $S$-matrix in these states, the infrared singularities which are contained in the S-matrix are canceled by additional ones coming from the interactions with the photons in the dressed states.  These matrix elements
 are finite. The statement is that
 $
 ~_D<\alpha|S^\mu|\beta>_D
   $ are have a finite limit as $\mu\to 0$.  This  moreover, the probabilities of transitions agree with those which are computed
   in the inclusive approach, 
   $$
    | ~_D<\alpha|S^\mu|\beta>_D|^2 = \sum_{\mu<\gamma<\lambda}| <\alpha|S^\mu |\beta,\gamma>|^2
     =|<\alpha|S^\lambda|\beta>|^2
     $$
     and the result is as if one simply computed the usual perturbative $S$ matrix for hard particles, but with the detector resolution $\lambda$ as an infrared cutoff for the otherwise infrared divergent internal loops in Feynman diagrams. 
     
Dressing is a canonical transformation.  However, when the fundamental cutoff is removed, the canonical transformation in
equations (\ref{dress1})-(\ref{dress2}) is not a proper unitary transformation.  Every undressed state in the undressed Hilbert space is orthogonal to every dressed state in the dressed Hilbert space.
This means that, if the photon were truly massless, the dressed and undressed formalisms are inequivalent quantizations of quantum electrodynamics.   

 What is more, there is a fundamental difference between the two procedures.  This difference appears on the off-diagonal elements of the density matrix.  With dressed states, the production of soft photons is already included in the state and there is no further soft photon production when charged particles scatter.  Pure states evolve to pure states and there is no decoherence.  In the inclusive formalism, as we have argued, there should be some fundamental decoherence and even suppression of some scattering.    These are, in principle, physical differences which could be measured by experiments.  
  
 When $\mu\to 0$, the dressed states have other peculiarities.  For example the dressed states are never eigenstates of the total momentum.  They are always mixtures of states with different momenta, the spread of momenta being governed by the detector resolution.  They are also not Lorentz invariant.  This is apparent in that the coherent photon field has a classical piece.  If the
 dressing were obtained by a unitary transformation, we could Lorentz transform a state simply by undressing it, Lorentz transforming it, and then re-dressing it.  When dressing and undressing by a proper unitary transformation is not possible, the Lorentz transform  is not a proper unitary transformation.  It is not clear what the implications of this subtlety are.  There has already been some discussion of it
 in the context of infrared divergences \cite{Frohlich:1979uu}-\cite{Balachandran:2013wsa} and it would be interesting to understand  that work in the present context.
 
   \section{Epilogue}
   
\begin{figure}
 ~~~~~~~~~~~~~~~~~~~~~~~~~~~~~~~ \includegraphics[scale=.30]{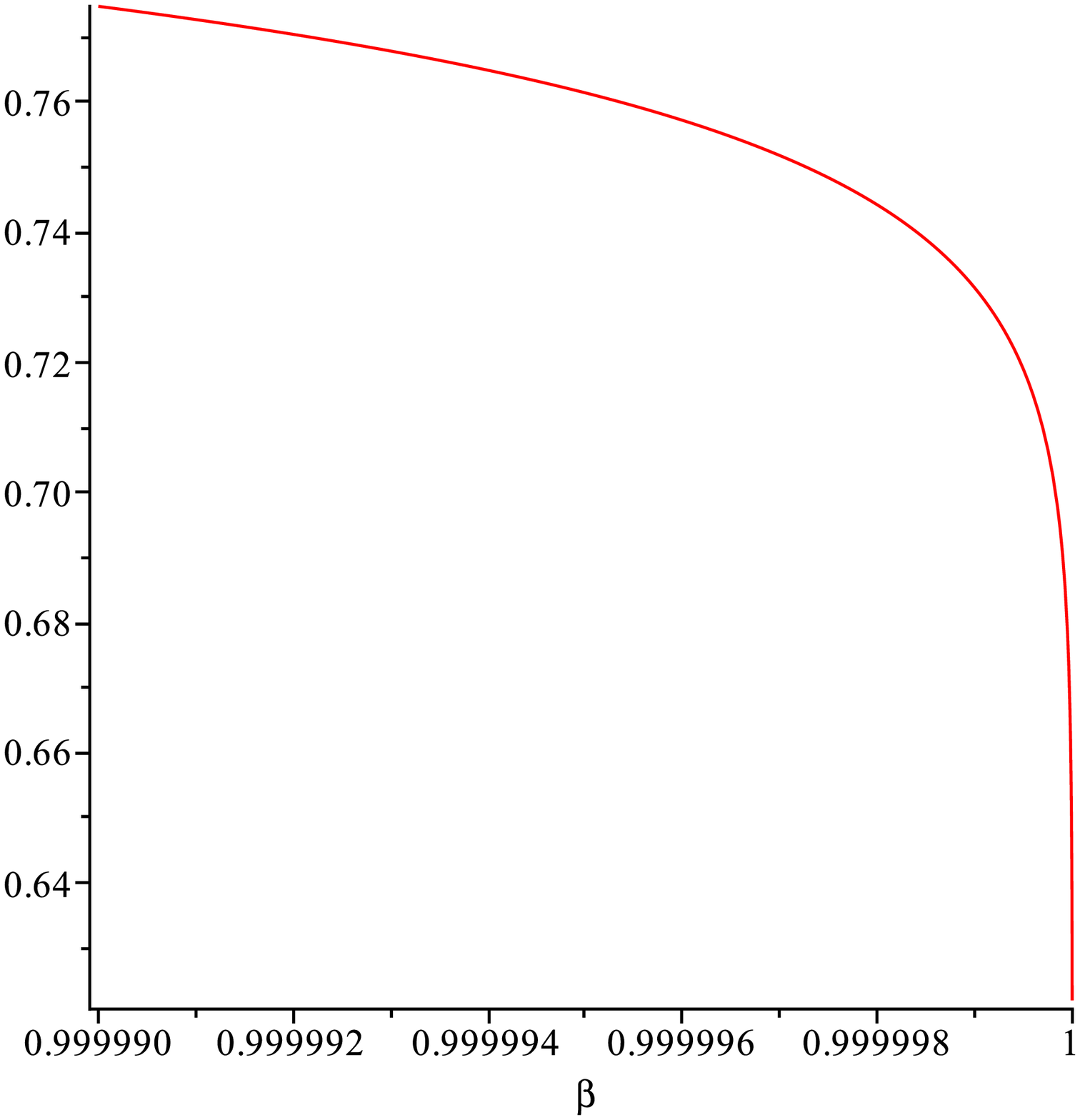}
\begin{caption}{The magnitude of the suppression factor for off-diagonal elements of the outgoing density matrix for Compton scattering when the time scale of the experiment is of the order of one second and the detector resolution is the electron mass. is  plotted on the vertical axis versus the relative velocity $\beta$ of the outgoing electrons  on the horizontal axis. }\end{caption}
\label{graph}\end{figure}

   We have argued that, in the limit where fundamental infrared cutoffs are removed, there are two fundamentally different interpretations of quantum electrodynamics and perturbative quantum gravity.  What is more, the differences are measurable in principle, although perhaps very difficult in practice.  For example, in a non-ideal scattering experiment, one which takes place over a finite time, a rough estimate of the decoherence effect would be to replace $\mu$ by the inverse time.  For example, if we consider Compton scattering, where the in-state is an electron and a hard photon and the out-state is also an electron and hard photon, the off-diagonal elements of the density matrix have the suppression factor
   $$
   \rho_{p,k;p',k'}\sim\left( \frac{\mu}{\lambda}\right)^{\frac{e^2}{4\pi^2}\left(\frac{1}{2\beta}\ln\frac{1+\beta}{1-\beta} -1\right)}
   $$
   where $\beta^2=1-\frac{m^4}{(p^\mu p_\mu')^2}$ is the relativistic relative velocity of the out-going electrons, with momenta $p,p'$, on the two legs of the reduced density matrix. If we take the detector resolution $\lambda$ to be the electron mass and $\mu$ to be an inverse second, the value of this suppression factor is graphed as a function of $\beta$ in figure \ref{graph}.      We see there that the suppression is significant only for very far off-diagonal elements where the relative velocity is close to that of light.   In fact, to get a suppression factor of 1/2, with one electron at rest, the other would need an energy of 100Gev.   
   For dressed states, one might worry about locality as the state is created by excitations which occupy far separated positions in space.  The breakdown of Lorentz invariance and the fact that states are not eigenstates of the momentum are also consequences that deserve  attention.  This balances the alternative of fundamental decoherence of the inclusive   approach.
   This fundamental decoherence is likely very small (and even smaller for perturbative quantum gravity) in any realistic interaction of charged particles. It would be interesting to find an experimental scenario where it would be detectable.  

\begin{acknowledgement}
The author acknowledges the collaboration of Dan Carney, Dominik Neuenfeld, Laurent Chaurette and Gianluca Grignani on most of the ideas and results that have been reported here. 
This work was supported by NSERC of Canada
\end{acknowledgement}
\end{document}